\documentclass[11pt]{article}
\usepackage{url}
\usepackage{booktabs}
\usepackage[round,sectionbib,sort&compress]{natbib}
\usepackage{caption}
\usepackage[french,english]{babel}

\usepackage{hyperref}
\usepackage[]{amsmath}
\usepackage[amsmath,thref,hyperref]{ntheorem}
\usepackage{cases}
\usepackage[]{amssymb}
\usepackage[]{amstext}
\usepackage[toc,page]{appendix}
\usepackage{amsxtra}
\usepackage[]{color}
\usepackage[]{eurosym}
\usepackage[]{float}
\usepackage[]{geometry}
\usepackage[]{graphicx}
\usepackage[]{capt-of}
\usepackage[]{indentfirst}
\usepackage[ansinew]{inputenc}
\usepackage[left,mathlines]{lineno}
\usepackage[]{lscape}
\usepackage[]{multicol}
\usepackage{multirow}
\usepackage[]{natbib}
\usepackage[]{parskip}
\usepackage{pgfplots}
    \pgfplotsset{compat=newest}
\usepackage[section]{placeins}
\usepackage[]{rotating}
\usepackage[onehalfspacing]{setspace}
\usepackage[]{subfig}
\usepackage{ifthen}
\usepackage{tikz}
\usepackage{tikz-3dplot}
\usetikzlibrary{petri,patterns}
\usepgflibrary{patterns}
\usepackage[bf,sf,compact,topmarks,small]{titlesec}
\usepackage[]{xcolor}
\usepackage{ulem}
\usepackage{datetime}
\usepackage{verbatim}
\usepackage{eqnarray}
\usepackage{etex}
\usepackage[neverdecrease]{paralist}
\usepackage{mdwlist}
\usepackage{pifont}
\newdate{date}{16}{08}{2016}
\usepackage{array}

\setdefaultleftmargin{3.5em}{2.2em}{1.87em}{1.7em}{1em}{1em}

\usepackage[french,english]{cleveref}
\usepackage{marginnote}
\theoremstyle{plain}

\newcounter{biscompt}

\theoremstyle{definition}

\newcounter{bisscompt}

\newtheorem{definition}{Definition}[section]

\theoremstyle{remark}

\newtheorem{conclusion}{Conclusion}
\theoremstyle{plain}
\theorembodyfont{\upshape}

\newtheorem*{acknowledgement}{Acknowledgement}

\crefformat{eqarray}{#2 Equation~#1#3}
\Crefformat{eqarray}{#2 Equations~#1#3}
\crefname{eqarray}{Equation}{Equations}
\crefmultiformat{eqarray}{#2 Equation~#1#3}%
{ and~#2#1#3}{, #2#1#3}{ and~#2#1#3}
\Crefmultiformat{eqarray}{#2 Equations~#1#3}%
{ and~#2#1#3}{, #2#1#3}{ and~#2#1#3}

\crefformat{condition}{#2 Condition~#1#3}
\Crefformat{condition}{#2 Conditions~#1#3}
\crefname{condition}{Condition}{Conditions}
\crefmultiformat{condition}{#2 Condition~#1#3}%
{ and~#2#1#3}{, #2#1#3}{ and~#2#1#3}
\Crefmultiformat{condition}{#2 Conditions~#1#3}%
{ and~#2#1#3}{, #2#1#3}{ and~#2#1#3}

\crefformat{bis}{#2 Condition~#1#3}
\Crefformat{bis}{#2 Conditions~#1#3}
\crefname{bis}{Condition}{Conditions}
\crefmultiformat{bis}{#2 Condition~#1#3}%
{ and~#2#1#3}{, #2#1#3}{ and~#2#1#3}
\Crefmultiformat{bis}{#2 Conditions~#1#3}%
{ and~#2#1#3}{, #2#1#3}{ and~#2#1#3}

\crefformat{theo}{#2 Condition~#1#3}
\Crefformat{theo}{#2 Conditions~#1#3}
\crefname{theo}{Condition}{Conditions}
\crefmultiformat{theo}{#2 Condition~#1#3}%
{ and~#2#1#3}{, #2#1#3}{ and~#2#1#3}
\Crefmultiformat{theo}{#2 Conditions~#1#3}%
{ and~#2#1#3}{, #2#1#3}{ and~#2#1#3}

\crefformat{proposition}{#2 Proposition~#1#3}
\Crefformat{proposition}{#2 Propositions~#1#3}
\crefname{proposition}{Proposition}{Propositions}
\crefmultiformat{proposition}{#2 Proposition~#1#3}%
{ and~#2#1#3}{, #2#1#3}{ and~#2#1#3}
\Crefmultiformat{proposition}{#2 Propositions~#1#3}%
{ and~#2#1#3}{, #2#1#3}{ and~#2#1#3}

\crefformat{corollary}{#2 Corollary~#1#3}
\Crefformat{corollary}{#2 Corollaries~#1#3}
\crefname{corollary}{Corollary}{Corollaries}
\crefmultiformat{corollary}{#2 Corollary~#1#3}%
{ and~#2#1#3}{, #2#1#3}{ and~#2#1#3}
\Crefmultiformat{corollary}{#2 Corollaries~#1#3}%
{ and~#2#1#3}{, #2#1#3}{ and~#2#1#3}

\crefformat{innerpropositionn}{#2 Proposition~#1#3}
\Crefformat{innerpropositionn}{#2 Propositions~#1#3}
\crefname{innerpropositionn}{Proposition}{Propositions}
\crefmultiformat{innerpropositionn}{#2 Proposition~#1#3}%
{ and~#2#1#3}{, #2#1#3}{ and~#2#1#3}
\Crefmultiformat{innerpropositionn}{#2 Propositions~#1#3}%
{ and~#2#1#3}{, #2#1#3}{ and~#2#1#3}

\crefformat{definition}{#2 Definition~#1#3}
\Crefformat{definition}{#2 Definitions~#1#3}
\crefname{definition}{Definition}{Definitions}
\crefmultiformat{definition}{#2 Definition~#1#3}%
{ and~#2#1#3}{, #2#1#3}{ and~#2#1#3}
\Crefmultiformat{definition}{#2 Definitions~#1#3}%
{ and~#2#1#3}{, #2#1#3}{ and~#2#1#3}

\crefrangeformat{definition}{#3 Definition~#1#4 to~#5#2#6}
\Crefrangeformat{definition}{#3 Definitions~#1#4 to~#5#2#6}

\crefformat{property}{#2 Property~#1#3}
\Crefformat{property}{#2 Properties~#1#3}
\crefname{property}{Property}{Properties}
\crefmultiformat{property}{#2 Property~#1#3}%
{ and~#2#1#3}{, #2#1#3}{ and~#2#1#3}
\Crefmultiformat{property}{#2 Properties~#1#3}%
{ and~#2#1#3}{, #2#1#3}{ and~#2#1#3}

\crefrangeformat{property}{#3 Property~#1#4 to~#5#2#6}
\Crefrangeformat{property}{#3 Properties~#1#4 to~#5#2#6}

\crefformat{innerpropertyy}{#2 Property~#1#3}
\Crefformat{innerpropertyy}{#2 Properties~#1#3}
\crefname{innerpropertyy}{Property}{Properties}
\crefmultiformat{innerpropertyy}{#2 Property~#1#3}%
{ and~#2#1#3}{, #2#1#3}{ and~#2#1#3}
\Crefmultiformat{innerpropertyy}{#2 Properties~#1#3}%
{ and~#2#1#3}{, #2#1#3}{ and~#2#1#3}

\crefrangeformat{innerpropertyy}{#3 Property~#1#4 to~#5#2#6}
\Crefrangeformat{innerpropertyy}{#3 Properties~#1#4 to~#5#2#6}

\crefformat{biss}{#2 Property~#1#3}
\Crefformat{biss}{#2 Properties~#1#3}
\crefname{biss}{Property}{Properties}
\crefmultiformat{biss}{#2 Property~#1#3}%
{ and~#2#1#3}{, #2#1#3}{ and~#2#1#3}
\Crefmultiformat{biss}{#2 Properties~#1#3}%
{ and~#2#1#3}{, #2#1#3}{ and~#2#1#3}

\crefrangeformat{biss}{#3 Property~#1#4 to~#5#2#6}
\Crefrangeformat{biss}{#3 Properties~#1#4 to~#5#2#6}

\crefformat{prop}{#2 Property~#1#3}
\Crefformat{prop}{#2 Properties~#1#3}
\crefname{prop}{Property}{Properties}
\crefmultiformat{prop}{#2 Property~#1#3}%
{ and~#2#1#3}{, #2#1#3}{ and~#2#1#3}
\Crefmultiformat{prop}{#2 Properties~#1#3}%
{ and~#2#1#3}{, #2#1#3}{ and~#2#1#3}

\crefrangeformat{prop}{#3 Property~#1#4 to~#5#2#6}
\Crefrangeformat{prop}{#3 Properties~#1#4 to~#5#2#6}

\crefformat{secinapp}{#2 Appendix~#1#3}
\Crefformat{secinapp}{#2 Appendices~#1#3}
\crefname{secinapp}{Appendix}{Appendices} 
\crefmultiformat{secinapp}{#2 Appendix~#1#3}%
{ and~#2#1#3}{, #2#1#3}{ and~#2#1#3}
\Crefmultiformat{secinapp}{#2 Appendices~#1#3}%
{ and~#2#1#3}{, #2#1#3}{ and~#2#1#3}

\crefformat{section}{#2 Section~#1#3}
\Crefformat{section}{#2 Sections~#1#3}
\crefname{section}{Section}{Sections}
\crefmultiformat{section}{#2 Section~#1#3}%
{ and~#2#1#3}{, #2#1#3}{ and~#2#1#3}
\Crefmultiformat{section}{#2 Sections~#1#3}%
{ and~#2#1#3}{, #2#1#3}{ and~#2#1#3}

\crefrangeformat{section}{#3 Section~#1#4 to~#5#2#6}
\Crefrangeformat{section}{#3 Sections~#1#4 to~#5#2#6}

\crefformat{subsubsection}{#2 Section~#1#3}
\Crefformat{subsubsection}{#2 Sections~#1#3}
\crefname{subsubsection}{Section}{Sections}
\crefmultiformat{subsubsection}{#2 Section~#1#3}%
{ and~#2#1#3}{, #2#1#3}{ and~#2#1#3}
\Crefmultiformat{subsubsection}{#2 Sections~#1#3}%
{ and~#2#1#3}{, #2#1#3}{ and~#2#1#3}

\crefrangeformat{subsubsection}{#3 Section~#1#4 to~#5#2#6}
\Crefrangeformat{subsubsection}{#3 Sections~#1#4 to~#5#2#6}

\crefformat{equation}{#2 Equation~#1#3}
\Crefformat{equation}{#2 Equations~#1#3}
\crefname{equation}{Equation}{Equations}
\crefmultiformat{equation}{#2 Equation~#1#3}%
{ and~#2#1#3}{, #2#1#3}{ and~#2#1#3}
\Crefmultiformat{equation}{#2 Equations~#1#3}%
{ and~#2#1#3}{, #2#1#3}{ and~#2#1#3}

\crefformat{Remark}{#2 Remark~#1#3}
\Crefformat{Remark}{#2 Remarks~#1#3}
\crefname{Remark}{Remark}{Remarks}
\crefmultiformat{Remark}{#2 Remark~#1#3}%
{ and~#2#1#3}{, #2#1#3}{ and~#2#1#3}
\Crefmultiformat{Remark}{#2 Remarks~#1#3}%
{ and~#2#1#3}{, #2#1#3}{ and~#2#1#3}

\crefformat{example}{#2 Example~#1#3}
\Crefformat{example}{#2 Examples~#1#3}
\crefname{example}{Example}{Examples}
\crefmultiformat{example}{#2 Example~#1#3}%
{ and~#2#1#3}{, #2#1#3}{ and~#2#1#3}
\Crefmultiformat{example}{#2 Examples~#1#3}%
{ and~#2#1#3}{, #2#1#3}{ and~#2#1#3}

\crefformat{definition}{#2 Definition~#1#3}
\Crefformat{definition}{#2 Definitions~#1#3}
\crefname{definition}{Definition}{Definitions}
\crefmultiformat{definition}{#2 Definition~#1#3}%
{ and~#2#1#3}{, #2#1#3}{ and~#2#1#3}
\Crefmultiformat{definition}{#2 Definitions~#1#3}%
{ and~#2#1#3}{, #2#1#3}{ and~#2#1#3}

\crefformat{table}{#2 Table~#1#3}
\Crefformat{table}{#2 Tables~#1#3}
\crefname{table}{Table}{Tables}
\crefmultiformat{table}{#2 Table~#1#3}%
{ and~#2#1#3}{, #2#1#3}{ and~#2#1#3}
\Crefmultiformat{table}{#2 Tables~#1#3}%
{ and~#2#1#3}{, #2#1#3}{ and~#2#1#3}

\crefformat{figure}{#2 Figure~#1#3}
\Crefformat{figure}{#2 Figures~#1#3}
\crefname{figure}{Table}{Figures}
\crefmultiformat{figure}{#2 Figure~#1#3}%
{ and~#2#1#3}{, #2#1#3}{ and~#2#1#3}
\Crefmultiformat{figure}{#2 Figures~#1#3}%
{ and~#2#1#3}{, #2#1#3}{ and~#2#1#3}

\crefformat{conclusion}{#2 Conclusion~#1#3}
\Crefformat{conclusion}{#2 Conclusions~#1#3}
\crefname{conclusion}{Conclusion}{Conclusions}
\crefmultiformat{conclusion}{#2 Conclusion~#1#3}%
{ and~#2#1#3}{, #2#1#3}{ and~#2#1#3}
\Crefmultiformat{conclusion}{#2 Conclusions~#1#3}%
{ and~#2#1#3}{, #2#1#3}{ and~#2#1#3}

\crefrangeformat{conclusion}{#3 Conclusion~#1#4 to~#5#2#6}
\Crefrangeformat{conclusion}{#3 Conclusions~#1#4 to~#5#2#6}

\crefformat{remark}{#2 Remark~#1#3}
\Crefformat{remark}{#2 Remarks~#1#3}
\crefname{remark}{Remark}{Remarks}
\crefmultiformat{remark}{#2 Remark~#1#3}%
{ and~#2#1#3}{, #2#1#3}{ and~#2#1#3}
\Crefmultiformat{remark}{#2 Remarks~#1#3}%
{ and~#2#1#3}{, #2#1#3}{ and~#2#1#3}

\crefrangeformat{remark}{#3 Remark~#1#4 to~#5#2#6}
\Crefrangeformat{remark}{#3 Remarks~#1#4 to~#5#2#6}

\hypersetup{
    bookmarks=true,         
    unicode=false,          
    pdftoolbar=true,        
    pdfmenubar=true,        
    pdffitwindow=false,     
    pdfstartview={FitH},    
    pdftitle={My title},    
    pdfauthor={Author},     
    pdfsubject={Subject},   
    pdfcreator={Creator},   
    pdfproducer={Producer}, 
    pdfkeywords={keyword1} {key2} {key3}, 
    pdfnewwindow=true,      
    colorlinks=true,       
    linkcolor=blue,          
    citecolor=blue,        
    filecolor=black,      
    urlcolor=blue           
}

\DeclareGraphicsExtensions{.png,.jpeg}

    \makeatletter
    \def\tagform@#1{\maketag@@@{\ignorespaces#1\unskip\@@italiccorr}}
    \renewcommand{\theequation}{(\oldtheequation)}
    \makeatother

 \title{\vspace{-1cm}\textbf{Risk reduction and Diversification within Markowitz's Mean-Variance Model: Theoretical Revisit}\footnote{Corresponding author: Gilles Boevi KOUMOU, \href{mailto:nettey-boevi-gilles-b.koumou.1@ulaval.ca}{nettey-boevi-gilles-b.koumou.1@ulaval.ca}}\\ \vspace{0mm}  {\Large \vspace{0.2cm}}} 

\author{Gilles Boevi KOUMOU\footnote{CIRP\'{E}E et D\'{e}partement d'\'{e}conomique, Universit\'{e} Laval, Qu\'{e}bec, Canada. Email: \href{mailto:nettey-boevi-gilles-b.koumou.1@ulaval.ca}{nettey-boevi-gilles-b.koumou.1@ulaval.ca}} 
 \vspace{0.5cm}
}

\date{\displaydate{date}}

\begin{document}
\maketitle

\thispagestyle{empty}
\pagenumbering{arabic}

\nolinenumbers

%
%
\begin{spacing}{1}

\begin{abstract}
\vspace{0.0cm}
\begin{footnotesize}
The conventional wisdom of mean-variance (MV) portfolio theory asserts that the nature of the relationship between risk and diversification is a decreasing asymptotic function, with the asymptote approximating the level of portfolio systematic risk or undiversifiable risk. This literature assumes that investors hold an equally-weighted or a MV portfolio and quantify portfolio diversification using portfolio size. However, the equally-weighted portfolio and portfolio size are MV optimal if and only if asset returns distribution is exchangeable or investors have no useful information about asset expected return and risk. Moreover, the whole of literature, absolutely all of it, focuses only on risky assets, ignoring the role of the risk free asset in the efficient diversification. Therefore, it becomes interesting and important to answer this question: how valid is this conventional wisdom when investors have full information about asset expected return and risk and asset returns distribution is not exchangeable in both the case where the risk free rate is available or not? Unfortunately,
this question have never been addressed in the current literature. This paper fills the gap.
\vspace{0.1cm}
\end{footnotesize}

\end{abstract}

\end{spacing}

%
%
\newpage
\section{Introduction}\label{intro}

The conventional wisdom of mean-variance (MV) portfolio theory asserts that the nature of the relationship between risk and diversification is a decreasing asymptotic function, with the asymptote approximating the level of portfolio systematic risk or undiversifiable risk \citep[see][]{EvansArcher1968,Wagner1971,EVANS1975,Elton1977,Statman1987,Bird1986,Lloyd1981,BLOOMFIELD1977,JOHNSON1974}. This literature assumes that investors hold an equally-weighted or a MV portfolio and quantify portfolio diversification using portfolio size. However, the equally-weighted portfolio and portfolio size are MV optimal if and only if asset returns distribution is exchangeable or investors have no useful information about asset expected return and risk \citep[see][]{Markowitz1952,Samuelson1967}. Moreover, the whole of literature, absolutely all of it, focuses only on risky assets, ignoring the role of the risk free asset in the efficient diversification. Therefore, it becomes interesting and important to answer this question: how valid is this conventional wisdom when investors have full information about asset expected return and risk and asset returns distribution is not exchangeable in both the case where the risk free rate is available or not? Unfortunately, this question have never been addressed in the current literature. 

The objective of this paper is to fill this gap. To do this, we examine analytically the nature of the relationship between risk and diversification in the MV model when investors have full information about asset expected return and risk and asset returns distribution is not exchangeable. We consider two cases, one in which assets are risky and one assuming the existence of the risk free asset. In both cases, we assume that short sales are allowed. We measure risk by the variance of a portfolio return as suggested in the MV model. We quantify diversification using a new measure other than portfolio size. We derive this measure exploiting the definition of the preference for diversification in the expected utility theory (EUT) and the relation between the EUT and the MV model. By doing so, we demonstrate, contrary to \citet{Fernholz2010}, that there is a specific measure of portfolio diversification in the MV model. This measure is the approximate version of \textquotedblleft diversification return\textquotedblright, a measure of portfolio diversification introduced by \citet{FamaBoth1992} and by \citet{Fernholz1982,Fernholz2010} under the name of excess growth in stochastic portfolio theory. Our results suggest, in both cases, that the conventional wisdom no longer holds, except for particular values of the MV model inputs. The nature of the relationship between risk and diversification is an inverted U-shaped concave function in the diversification-risk plane.





The rest of the paper is organized as follows. In \cref{DM}, we introduce the new portfolio diversification measure considered in this paper. In \cref{RD}, we derive the analytical relationship between risk and diversification. \cref{conclusion} concludes.

 
\section{Measuring Diversification}\label{DM}
The essential of the existing literature measures portfolio diversification using portfolio size. This measure was introduced by \citet{EvansArcher1968} inspiring from the capital asset pricing model (CAPM). However, while portfolio size is an optimal portfolio diversification measure when asset returns distribution is exchangeable or investors have no useful information about asset expected return and risk, this measure
is insufficient to accurately quantify portfolio diversification when investors have full information and asset returns distribution is not exchangeable. For example, even before \citet{EvansArcher1968}'s work, \citet{Markowitz1952} reports that:  \begin{quote}\textquotedblleft \textit{The adequacy 
of diversification is not thought by investors to depend solely on the 
number of different securities held.}\textquotedblright
\end{quote}
\citet{Sharpe1972} also reports that:  
\begin{quote}
\textquotedblleft \textit{The number of securities in a portfolio provides a fairly crude measure of diversification}\textquotedblright
\end{quote}
Therefore, in this paper, we propose a new measure to quantify portfolio diversification. We derive this measure exploiting the definition of the preference for diversification (PFD) in the expected utility theory (EUT) and the relation between the EUT and the mean-variance (MV) model.

There are several notions of PFD in theory of choice under uncertainty, but all are proven to be equivalent in the EUT \citep[see][]{Chateauneuf2002}. Thus, we focus only on that defined in \citet{Dekel1989} and extended later by \citet{Chateauneuf2002,Chateauneuf2007} to the space of random variables, because it facilitates our analysis. Consider a decision maker who chooses from $\mathcal{R}$, a vector space 
of bounded real-valued random variables. For $R \in \mathcal{R}$, $F_R$ will denote the cumulative distribution function of $R$. Let $\succeq$ be the preference relation over $\mathcal{R}$ of a decision maker. Assume that $\succeq$ has a von Neumann-Morgenstern (VNM) expected utility representation i.e. there exists a function $u:\mathcal{R} \longrightarrow \mathbb{R}$ such that
\begin{equation*}
R_1 \succeq R_2 \Longleftrightarrow \mathbb{E}u(R_1)\geq \mathbb{E}u(R_2),
\end{equation*}
where $\mathbb{E}u(R)=\int u(r)dF_R(r)$. Moreover, $u$ is unique up to positive affine transformations.

\begin{definition}[\citet{Chateauneuf2002}]\label{defpd} The preference relation $\succeq$ exhibits preference for diversification if for any $R_i,\,i=1,...,N \in \mathcal{R}$ and $w_i \in [0,1],\,\,i=1,...,N$ such that $\sum_{i=1}^{N}w_i=1$,
\begin{equation*}
R_1\sim R_2\sim...\sim R_N\Rightarrow \sum_{i=1}^{N}w_i\,R_i\succeq R_j \,\,\,\,\,\forall \,j=1,...,N.
\end{equation*}
\end{definition}
\cref{defpd} implies that
\begin{equation*}
\mathbb{E}u(R_1)=...=\mathbb{E}u(R_N)\Longrightarrow \mathbb{E}u\left(\sum_{i=1}^{N}w_i\,R_i\right)\geq \mathbb{E}u(R_j) \,\,\,\,\,\forall \,j=1,...,N.
\end{equation*}
The gain of diversification in the EUT can therefore be measured by 
\begin{equation*}
\underline{\mathbb{E}u}\left(\boldsymbol{w}\right)=\mathbb{E}u\left(\sum_{i=1}^{N}w_i\,R_i\right)- \mathbb{E}u(R_j),
\end{equation*}
where $\mathbb{E}u(R_1)=...=\mathbb{E}u(R_N)$ and $\boldsymbol{w}=(w_1,...,w_N)^{\top}$. By analogy, the gain of diversification in the EUT can be measured in general by 
\begin{equation*}
\underline{\mathbb{E}u}\left(\boldsymbol{w}\right)=\mathbb{E}u\left(\sum_{i=1}^{N}w_i\,R_i\right)- \sum_{i=1}^{N}w_i\,\mathbb{E}u(R_i).
\end{equation*}

It is well-known that the MV model is a special case of the EUT when utility function is the exponential utility function and asset returns are elliptical distributed. Therefore, the definition of the PFD in the EUT is also valid for the MV model, since it is invariant to asset returns distribution. It follows that the benefit of diversification in the MV model can be measured by
\begin{equation}
\underline{U}_{MV}\left(\boldsymbol{w}\right)=U_{MV}\left(\sum_{i=1}^{N}w_i\,R_i\right)- \sum_{i=1}^{N}w_i\,U_{MV}(R_i),\label{eqdm}
\end{equation}
 where $U_{MV}(.)$ is the MV utility function 
 \begin{equation}
U_{MV}(R)=\mathbb{E}(R)-\frac{\gamma}{2}\,\mathbb{V}(R)\label{eqmv}
\end{equation}
with $\gamma$ is the investor's risk aversion coefficient, $\mathbb{E}(.)$ the expectation operator and $\mathbb{V}(.)$ that of the variance. Substituting $U_{MV}(.)$ from \ref{eqmv} into \ref{eqdm}, we obtain
\begin{equation*}
\underline{U}_{MV}\left(\boldsymbol{w}\right)=\frac{\gamma}{2} \,\left(\sum_{i=1}^{N}w_i\,\sigma_i^2-\sigma^2(\boldsymbol{w})\right),\label{}
\end{equation*}
where $\sigma_i^2$ is the variance of asset $i$ and $\sigma^2(\boldsymbol{w})=\boldsymbol{w}^{\top}\boldsymbol{\Sigma}\,\boldsymbol{w}$ is portfolio variance with $\boldsymbol{\Sigma}$ the covariance matrix of asset returns. The portfolio diversification measure in the MV model is therefore
\begin{equation}
EDM(\boldsymbol{w})=\sum_{i=1}^{N}w_i\,\sigma_i^2-\sigma^2(\boldsymbol{w}).\label{eq3}
\end{equation}
As a result, contrary to \citet{Fernholz2010}, there is a specific measure of portfolio diversification in the MV model, which is the approximate version of \textquotedblleft diversification return\textquotedblright,~ a measure of portfolio diversification introduced by \citet{FamaBoth1992} and by \citet{Fernholz1982,Fernholz2010} under the name of \textquotedblleft excess growth rate\textquotedblright~ in stochastic portfolio theory. $EDM(\boldsymbol{w})$ can also be rewritten as follows
\begin{equation}
EDM(\boldsymbol{w})=\sum_{i=1}^{N}w_i\,\left(\sigma_i^2-\sigma^2(\boldsymbol{w})\right).\label{eq4}
\end{equation}
The term in the parenthesis, $\sigma_i^2-\sigma^2(\boldsymbol{w})$, measures the benefit of diversification, in term of risk reduction, to hold portfolio $w$ instead of to concentrate on asset $i$. It follows that $EDM(\boldsymbol{w})$ measures the average benefit of diversification, in term of risk reduction, to hold portfolio $\boldsymbol{w}$ instead of holding a single asset portfolio.

\section{Risk and diversification: analytical relationship}\label{RD}

In this section, we revisit the relationship between risk and diversification in the MV model. We consider two cases, one in which assets are risky and one assuming the existence of the risk free asset. We quantify the diversification by our proposed measure and the risk by the variance of a portfolio returns as recommended in the MV model. We present the relationship in the diversification-risk plane instead of the risk-diversification plane as usual. This facilitates our analysis. We assume that investors have full information and asset returns distribution is not exchangeable. In other words, investors hold the MV optimal portfolio and this portfolio is different to the equally-weighted portfolio. Further, we assume that short sales are unrestricted. This allows us to have an exact relationship. 


\subsection{Assets are risky}
Consider the case where we have $N$ risky assets. In that case, the MV optimal portfolio denoted by $\boldsymbol{w}^{MV}$ is 
\begin{equation}
\boldsymbol{w}^{MV}\equiv \boldsymbol{w}^{MV}(\tau)=\tau\,\left(\boldsymbol{\Sigma}^{-1}\boldsymbol{\mu} -\frac{B}{C}\boldsymbol{\Sigma}^{-1}\boldsymbol{1}\right)+\frac{\boldsymbol{\Sigma}^{-1}\boldsymbol{1}}{C},\label{eq5}
\end{equation} 
where $\tau=\frac{1}{\gamma}$ is investor's risk tolerance coefficient, $\boldsymbol{\mu}=(\mu_1,...,\mu_N)^{\top}$ is the vector of asset expected returns, $\boldsymbol{1}=(1,...,1)^{\top}$ is the vector column of ones, $B=\boldsymbol{\mu}^{\top}\boldsymbol{\Sigma}^{-1}\,\boldsymbol{1}$ and $C=\boldsymbol{1}^{ \top}\boldsymbol{\Sigma}^{-1}\,\boldsymbol{1}$. The risk of portfolio $\boldsymbol{w}^{MV}$, measured by its variance denoted by $\sigma^2\left(\boldsymbol{w}^{MV}\right)$, is
\begin{equation}
\sigma^2\left(\boldsymbol{w}^{MV}\right) \equiv \sigma^2 \circ \boldsymbol{w}^{MV}(\tau)=\frac{D}{C}\tau^2+\frac{1}{C},\label{eq6}
\end{equation}
where $D=AC-B^2$ with $A=\boldsymbol{\mu}^{\top}\boldsymbol{\Sigma}^{-1}\,\boldsymbol{\mu}$. As we can observe, portfolio risk is an increasing function of investor's risk tolerance coefficient $\tau$, since $\tau\geq 0$. 
We focus therefore on the relationship between investor's risk tolerance coefficient and diversification. The relationship between risk and diversification is presented in the \ref{other}.

From \ref{eq3} and \ref{eq5}, the diversification degree of portfolio $\boldsymbol{w}^{MV}$ is
\begin{equation}
EDM\left(\boldsymbol{w}^{MV}\right) \equiv EDM \circ \boldsymbol{w}^{MV}\left(\tau\right)=-\frac{D}{C}\tau^2+\frac{\left(EC-FB\right)}{C}\tau+\frac{F-1}{C},\label{eq7}
\end{equation}
where $E=\boldsymbol{\mu}^{\top}\boldsymbol{\Sigma}^{-1}\boldsymbol{\sigma}^2$ and $F=\boldsymbol{1}^{\top}\boldsymbol{\Sigma}^{-1}\boldsymbol{\sigma}^2$ with $\boldsymbol{\sigma}^2=(\sigma^2_1,...,\sigma^2_N)^{\top}$ the vector of asset variances. 
\cref{eq7} represents the relationship between investor's risk tolerance coefficient and diversification in the MV model when assets are risky, short sales are allowed, investors hold the MV optimal portfolio and asset returns distribution is not exchangeable. Examination of the first and second derivatives in respect of $\tau$ shows that $EDM\left(\boldsymbol{w}^{MV}\right)$ is strictly concave function of $\tau$ with a unique maximum point where $\frac{\partial EDM\left(\boldsymbol{w}^{MV}\right)}{\partial \tau}=0$, i.e.
   \begin{align}
\frac{\partial EDM\left(\boldsymbol{w}^{MV}\right)}{\partial \tau}=&-2\frac{D}{C}\tau+\frac{\left(EC-FB\right)}{C} \\
=&0 \text{ when } \tau=\frac{EC-FB}{2\,D}\\\label{eq21}
\frac{\partial^2 EDM\left(\boldsymbol{w}^{MV}\right)}{\partial \tau^2}&=-2\frac{D}{C} < 0
\end{align}
Since $\tau>0$, there are two cases in point depending of sign of $EC-FB$. Assume that $EC-FB$ has negative sign i.e. $EC-FB\leq0$. In that case, since $D>0$ and $C>0$,
\begin{equation}
\frac{\partial EDM\left(\boldsymbol{w}^{MV}\right)}{\partial \tau}<0.
\end{equation}
It follows that $EDM\left(\boldsymbol{w}^{MV}\right)$ is a strictly decreasing concave function of $\tau$ \citetext{see \cref{f1}}. An investor with a high (low) risk tolerance coefficient may hold a less (more) diversified portfolio. In other words, an investor with a high (low) risk aversion coefficient may hold a more (less) diversified portfolio.

\pgfplotsset{my personal style/.style=
{font=\footnotesize},width=8cm,height=7cm}
\begin{figure}[!t]
\caption{Risk and diversification relationship when investors have full information and asset returns distribution is not exchangeable: case where assets are risky}
\label{f1}
\begin{center}
\begin{tikzpicture}[]
\begin{axis}[my personal style,title=$EC-FB>0$,
xlabel=Risk ($\tau$),
ymax=,
ymin=-20,
xmin=,
ylabel=Efficient Diversification ($EMD$),legend pos=north west,legend cell align=left,legend style={draw=none}]
\addplot+[mark=none,domain=0:40,smooth,black] {-0.3324407*x^2+9.670447*x+39.01232};
\end{axis}
\end{tikzpicture}
\begin{tikzpicture}[]
\begin{axis}[my personal style,title=$EC-FB\leq0$,
xlabel=Risk ($\tau$),
ymax=,
ymin=-20,
xmin=,
ylabel=,legend pos=north west,legend cell align=left,legend style={draw=none}]
\addplot+[mark=none,domain=0:40,smooth,black] {-0.002524025*x^2-0.7829712*x+ 39.01232};
\end{axis}
\end{tikzpicture}
\begin{minipage}{12cm}%
  \begin{spacing}{0.5}
    \footnotesize
We consider a universe of four assets. The variance-covariances matrix is $\boldsymbol{\Sigma}=\begin{pmatrix} 
185.0 & 86.5 &  80 & 20.0\\
 86.5 &196.0  & 76 & 13.5\\
80.0 & 76.0 & 411 &-19.0
\\
20.0 & 13.5 & -19 & 25.0
\end{pmatrix}$. In the case where $EC-FB>0$, $\boldsymbol{\mu}=\begin{pmatrix} 
14& 12& 15& 7
\end{pmatrix}^{\top}$. Otherwise i.e. $EC-FB\leq0$ $\boldsymbol{\mu}=\begin{pmatrix} 
0.14& 0.12& 0.15& 0.7
\end{pmatrix}^{\top}$.
\end{spacing}
  \end{minipage}
\end{center}
\end{figure}
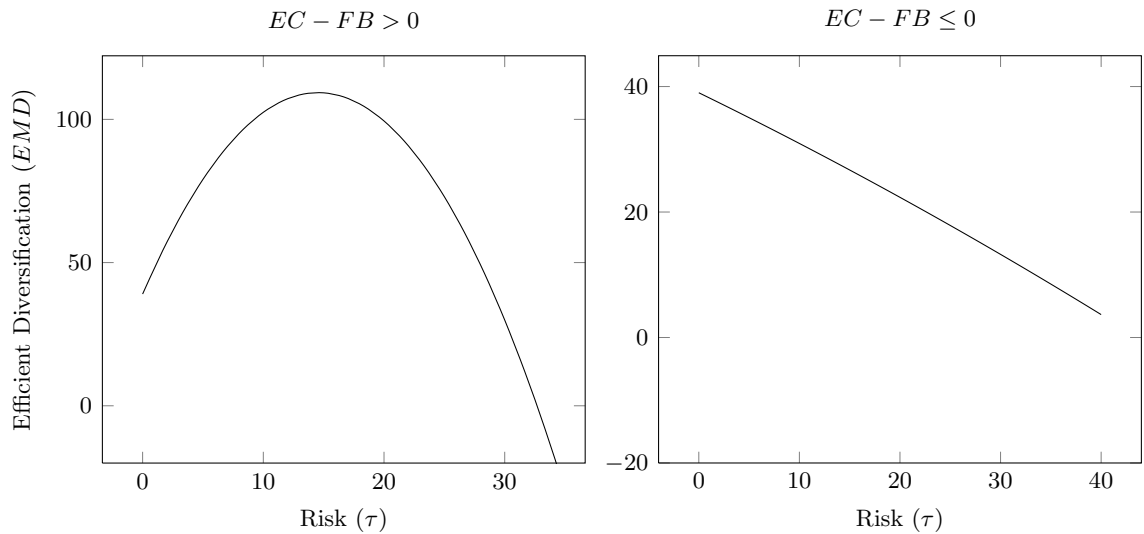

Now, assume that $EC-FB$ has positive sign i.e.  $EC-FB>0$. In that case, 
\begin{equation}
\begin{cases} \frac{ \partial EDM\left(\boldsymbol{w}^{MV}\right)}{\partial \tau}\geq 0 &\mbox{if } 0\leq\tau\leq \frac{EC-FB}{2\,D} \\ 
\frac{ \partial EDM\left(\boldsymbol{w}^{MV}\right)}{\partial \tau}\leq 0 &\mbox{if } \tau\geq \frac{EC-FB}{2\,D}. \end{cases}
\end{equation}
It follows that $EDM\left(\boldsymbol{w}^{MV}\right)$ is an inverted U-shaped concave function with maximum at $\tau=\frac{EC-FB}{2\,D}$ \citetext{see \cref{f1}}. The diversification is more attractive for investors with intermediate levels of risk aversion coefficient. The very risk averse investors should choose to hold a less diversified portfolios constituting mainly of less risky assets. Similarly, the very lower risk averse investors should choose to hold a less diversified portfolios, but consisting mainly of assets with higher expected return.


Since portfolio risk is an increasing function of investor's risk tolerance coefficient $\tau$, we have the following result.
\begin{conclusion} When investors have full information and asset returns distribution is not exchangeable, the conventional wisdom holds if only if the mean-variance inputs is defined such as $EC-FB\leq0$. Otherwise i.e. $EC-FB>0$, the conventional wisdom no longer holds. The nature of the relationship between risk and diversification is an inverted U-shaped concave function in the diversification-risk plane \citetext{see \cref{f3}}. A lower risk portfolio does not necessary exhibits higher diversification degree. The diversification is more attractive for investors with intermediate levels of risk aversion coefficient than investors with higher or lower levels of risk aversion coefficient.
\end{conclusion}


\subsection{Risk free is available}
Now, consider the case where the risk free asset is available. In that case, the MV optimal portfolio is a combination of the risk free asset and the tangent portfolio (risky portfolio) as follows $\boldsymbol{w}^{MV}=(w_f,(1-w_f)\boldsymbol{w}^{tg})$, where $w_f$ is the weight of the risk free asset and $\boldsymbol{w}^{tg}$ that of the tangent portfolio. The tangent portfolio $\boldsymbol{w}^{tg}$ is obtained independently of the risk aversion coefficient
\begin{equation}
\boldsymbol{w}^{tg} =\frac{\Sigma^{-1}(\boldsymbol{\mu}-\mu_f\,\boldsymbol{1})}{B-C\,\mu_f},\label{eq14}
\end{equation}
where $\mu_f$ is the risk free asset rate. 
While the risk free portfolio $w_f$ is obtained dependently of $\tau$
\begin{equation}
w_f \equiv \boldsymbol{w}(\tau)=1-\left(B-C\,\mu_f\right)\tau.\label{eq14}
\end{equation}
The risk of the tangent portfolio $w^{tg}$ is 
\begin{equation}
\sigma^2\left(\boldsymbol{w}^{tg}\right)=\frac{S^2}{(B-C\,\mu_f)^2}, \label{eq15} 
\end{equation}
where $S=\sqrt{C\mu_f^2-2B\mu_f+A}$. The risk of optimal portfolio $\boldsymbol{w}^{MV}$ is
\begin{equation}
\sigma^2\left(\boldsymbol{w}^{MV}\right)=(1-w_f)^2\sigma^2\left(\boldsymbol{w}^{tg}\right). \label{eq16}
\end{equation}
From \ref{eq14}, we have
\begin{equation}
1-w_f=\left(B-C\,\mu_f\right)\tau.\label{eq26}
\end{equation}
Substituting $1-w_f$ from \ref{eq26} into \ref{eq16}, we can write portfolio risk, $\sigma^2\left(\boldsymbol{w}^{MV}\right)$, as function of investor's risk tolerance coefficient $\tau$
\begin{equation}
\sigma^2\left(\boldsymbol{w}^{MV}\right)=\tau^2\left(B-C\,\mu_f\right)^2\sigma^2\left(\boldsymbol{w}^{tg}\right). \label{}
\end{equation}
In the existing literature, only the risk of tangent portfolio is considered. Doing so the existing literature ignores the role of the risk free asset in the efficient diversification. However, investors care about the risk of portfolio as whole, not only about the risk of the tangent portfolio. Further, risk free asset plays an important role in portfolio diversification, since its return is uncorrelated with risky assets. Thus, study the relationship between risk ($\sigma^2\left(\boldsymbol{w}^{MV}\right)$) and diversification ($EDM\left(\boldsymbol{w}^{MV}\right)$) is equivalent to study the relation between diversification and the weight of risk free asset ($w_f$) or investor's risk tolerance coefficient ($\tau$). We focus therefore on the relationship between $\tau$ and $EDM\left(\boldsymbol{w}^{MV}\right)$.  The relationship between risk and diversification is presented in \ref{other}. 

From \ref{eq3}, we have 
\begin{equation}
EDM\left(\boldsymbol{w}^{MV}\right)=(1-w_f)\sum_{i=1}^{N}w^{tg}_i\,\sigma_i^2-(1-w_f)^2\sigma^2\left(\boldsymbol{w}^{tg}\right).\label{eq27}
\end{equation}
Substituting $1-w_f$ from \ref{eq26} into \ref{eq27}, we write the equation for the diversification as a function of risk aversion coefficient as 
\begin{equation}
EDM\left(\boldsymbol{w}^{MV}\right)=\tau\,\left(B-C\,\mu_f\right)\sum_{i=1}^{N}w^{tg}_i\,\sigma_i^2-\tau^2\,(B-C\,\mu_f)^2\sigma^2(\boldsymbol{w}^{tg}).\label{eq28}
\end{equation}
Substituting $\boldsymbol{w}^{tg}$ from \ref{eq14} and $\sigma^2(\boldsymbol{w}^{tg})$ from \ref{eq15} into \ref{eq28}, we have
\begin{equation}
EDM\left(\boldsymbol{w}^{MV}\right)=\tau\,\left(E-F\,\mu_f\right)-\tau^2\,S^2.\label{eq29}
\end{equation}
\cref{eq29} represents the relationship between investor's risk tolerance coefficient and diversification in the MV model when risk free asset is available, short sales are allowed, investors hold the MV optimal portfolio and asset returns distribution is not exchangeable.
The first derivation of \ref{eq29} in respect of $\tau$ gives
\begin{equation}
\frac{\partial EDM\left(\boldsymbol{w}^{MV}\right)}{\partial \tau}=\left(E-F\,\mu_f\right)-2\tau\,S^2.
\end{equation}
The second derivation of \ref{eq29} in respect of $\tau$ gives
\begin{equation}
\frac{\partial^2 EDM\left(\boldsymbol{w}^{MV}\right)}{\partial \tau^2}=-2S^2.
\end{equation}
As in the case of risk assets, there are two cases in point depending of sign of $E-F\,\mu_f$:
\begin{equation}
\begin{cases} \frac{\partial EDM\left(\boldsymbol{w}^{MV}\right)}{\partial \tau}<0 &\mbox{if } E-F\,\mu_f\leq 0\\ 
\begin{cases} \frac{ \partial EDM\left(\boldsymbol{w}^{MV}\right)}{\partial \tau}\geq 0 &\mbox{if } \tau\leq \frac{2\,S^2}{E-F\,\mu_f} \\ 
\frac{ \partial EDM\left(\boldsymbol{w}^{MV}\right)}{\partial \tau}\leq 0 &\mbox{if } \tau\geq \frac{2\,S^2}{E-F\,\mu_f}. \end{cases} &\mbox{if } E-F\,\mu_f\geq 0. \end{cases}\label{eq244}
\end{equation}

\pgfplotsset{my personal style/.style=
{font=\footnotesize},width=7.5cm,height=7cm}
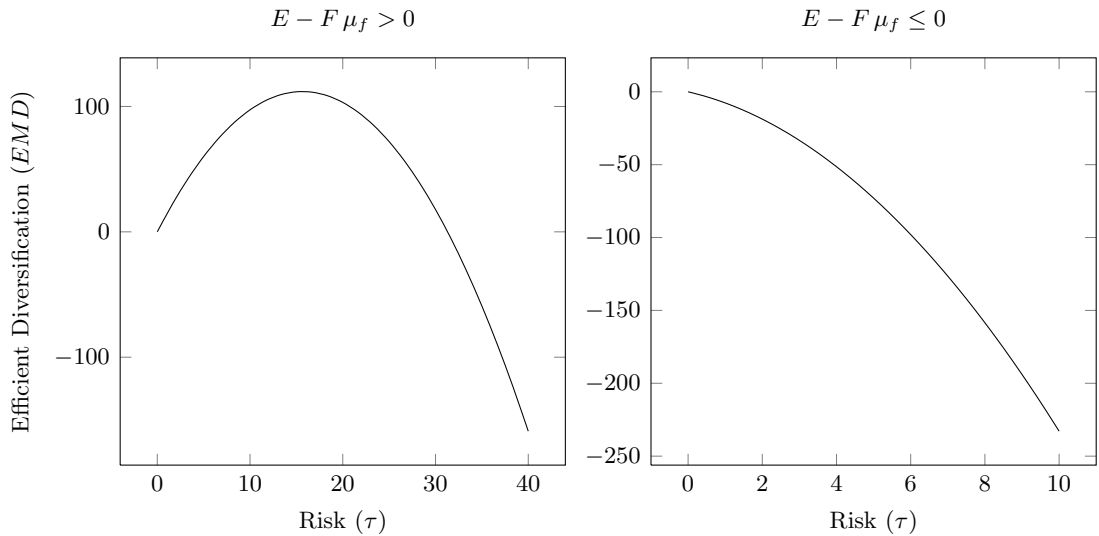
\begin{figure}[!t]
\caption{Risk and diversification relationship when investors have full information and asset returns distribution is not exchangeable: case where assets are risky}
\label{f2}
\begin{center}
\begin{tikzpicture}[]
\begin{axis}[my personal style,title=$E-F\,\mu_f>0$,
xlabel=Risk ($\tau$),
ymax=,
ymin=,
xmin=,
ylabel=Efficient Diversification ($EMD$),legend pos=north west,legend cell align=left,legend style={draw=none}]
\addplot+[mark=none,domain=0:40,smooth,black] {-0.6759361*0.6759361*x^2+14.30059*x};
\end{axis}
\end{tikzpicture}
\begin{tikzpicture}[]
\begin{axis}[my personal style,title=$E-F\,\mu_f\leq0$,
xlabel=Risk ($\tau$),
ymax=,
ymin=,
xmin=,
ylabel=,legend pos=north west,legend cell align=left,legend style={draw=none}]
\addplot+[mark=none,domain=0:10,smooth,black] {-1.318732*1.318732*x^2-5.89587*x};
\end{axis}
\end{tikzpicture}
\begin{minipage}{12cm}%
  \begin{spacing}{0.5}
    \footnotesize
We consider a universe of four assets. The variance-covariances matrix is $\boldsymbol{\Sigma}=\begin{pmatrix} 
185.0 & 86.5 &  80 & 20.0\\
 86.5 &196.0  & 76 & 13.5\\
80.0 & 76.0 & 411 &-19.0
\\
20.0 & 13.5 & -19 & 25.0
\end{pmatrix}$ and $\boldsymbol{\mu}=\begin{pmatrix} 
14& 12& 15& 7
\end{pmatrix}^{\top}$ and the risk free asset rate is $\mu_f=6$. In the case where $E-F\,\mu_f>0$, the risk free asset rate is $\mu_f=6$. Otherwise i.e. $E-F\,\mu_f\leq0$, the risk free asset rate is $\mu_f=13$.
\end{spacing}
  \end{minipage}
\end{center}
\end{figure}
From\ref{eq244}, we have the following result.
\begin{conclusion} When investors have full information, asset returns distribution is not exchangeable and risk free asset is available, the conventional wisdom holds if only if the mean-variance inputs is defined such as $E-F\,\mu_f\leq0$. Otherwise i.e. $E-F\,\mu_f>0$, the conventional wisdom no longer holds. The relationship between risk and diversification is an inverted U-shaped concave function in the diversification-risk plane \citetext{see \cref{f4}}. A lower risk portfolio does not necessary exhibits higher diversification degree. The diversification is more attractive for investors with intermediate levels of risk aversion coefficient than investors with higher or lower levels of risk aversion coefficient. 
\end{conclusion}

%
%

\section{Conclusion}\label{conclusion}
In this paper, we have revisited theoretically the nature of the relationship between risk and diversification in the mean-variance model. We have shown, regardless of whether or not the risk free asset is available, that the conventional wisdom, which asserts that the nature of the relationship between risk and diversification is a decreasing asymptotic function, with the asymptote approximating the level of portfolio systematic risk or undiversifiable risk, no longer holds when investors have full information about asset expected return and risk and asset returns distribution is not exchangeable regardless of whether or not the risk free asset is available, except for particular values of the MV model inputs. The nature of the relationship between risk and diversification is an inverted U-shaped concave function in the diversification-risk plane.

%
%

\begin{acknowledgement}
We gratefully acknowledge the financial
support from FQRSC.
\end{acknowledgement}

%
%

\newpage
\renewcommand\appendix{\par
  \setcounter{section}{0}
  \setcounter{subsection}{0}
  \renewcommand\thesection{Appendix \Alph{section}}
  \renewcommand\thesubsection{\Alph{section}.\arabic{subsection}}
} 
 \renewcommand{\theequation}{(\Alph{section}-\arabic{equation})}    
  \setcounter{equation}{0}
  
\appendix
\section{Risk and diversification: explicit relationship}\label{other}
Here, we derive the explicit relationship between risk and diversification in the risk-diversification plane. 
\subsection{Assets are risky}
From \ref{eq6}
we have 
\begin{equation}
\tau=\sqrt{\frac{C\,\sigma^2\left(\boldsymbol{w}^{MV}\right)-1}{D}}.\label{a1}
\end{equation}
Substituting for $\tau$ from \ref{a1} into \ref{eq7}, we write the equation for the diversification of a MV optimal portfolio as a function of its variance, as 
\begin{equation}
EDM\left(\boldsymbol{w}^{MV}\right)=\left(\frac{EC-FB}{C}\right)\sqrt{\frac{C\,\sigma^2\left(\boldsymbol{w}^{MV}\right)-1}{D}}-\sigma^2\left(\boldsymbol{w}^{MV}\right)+\frac{F}{C}\label{a3}
\end{equation}
\cref{a3} represents the relationship between risk and diversification in the MV model when assets are risky, short sales are allowed, investors hold the MV optimal portfolio and asset returns distribution is not exchangeable. The first partial derivative of $EDM\left(\boldsymbol{w}^{MV}\right)$ with respect of $\sigma^2\left(\boldsymbol{w}^{MV}\right)$ is
\begin{equation}
\frac{\partial EDM\left(\boldsymbol{w}^{MV}\right)}{\partial \sigma^2\left(\boldsymbol{w}^{MV}\right)}=\left(\frac{EC-FB}{C\sqrt{D}}\right)\frac{C}{2\sqrt{C\,\sigma^2\left(\boldsymbol{w}^{MV}\right)-1}}-1.
\end{equation}
Its second partial derivative with respect of $\sigma^2\left(\boldsymbol{w}^{MV}\right)$ is
\begin{equation}
\frac{\partial^2 EDM\left(\boldsymbol{w}^{MV}\right)}{\partial \sigma^2\left(\boldsymbol{w}^{MV}\right)^2}=-\frac{C}{4}\left(\frac{EC-FB}{C\sqrt{D}}\right)\left(C\,\sigma^2\left(\boldsymbol{w}^{MV}\right)-1\right)^{-3/2}
\end{equation}
There are two cases in point depending of sign of $EC-FB$. Assume that $EC-FB$ has negative sign i.e.  $EC-FB\leq0$. In that case, since $D>0$ and $C>0$,
\begin{align}
\frac{\partial EDM\left(\boldsymbol{w}^{MV}\right)}{\partial \sigma^2\left(\boldsymbol{w}^{MV}\right)}&<0\\
\frac{\partial^2 EDM\left(\boldsymbol{w}^{MV}\right)}{\partial \sigma^2\left(\boldsymbol{w}^{MV}\right)^2}&>0
\end{align}
It follows that diversification is an decreasing convex function of risk.  

Now, assume that $EC-FB$ has positive sign i.e.  $EC-FB>0$. In that case, 
\begin{equation}
\begin{cases} \frac{ \partial EDM\left(\boldsymbol{w}^{MV}\right)}{\partial \sigma^2\left(\boldsymbol{w}^{MV}\right)}\geq 0 &\mbox{if } \frac{1}{C}\leq\sigma^2\left(\boldsymbol{w}^{MV}\right)\leq \frac{1}{C}+\frac{(EC-FB)^2}{4DC} \\ 
\frac{ \partial EDM\left(\boldsymbol{w}^{MV}\right)}{\partial \sigma^2\left(\boldsymbol{w}^{MV}\right)}\leq 0 &\mbox{if } \sigma^2\left(\boldsymbol{w}^{MV}\right)\geq \frac{1}{C}+\frac{(EC-FB)^2}{4DC}. \end{cases}
\end{equation}
It follows that diversification is an inverted U-shaped concave function of risk with maximum when risk is equal to $\frac{1}{C}+\frac{(EC-FB)^2}{4DC}$ (see \cref{f3}).

\pgfplotsset{my personal style/.style=
{font=\footnotesize},width=7.5cm,height=7cm}
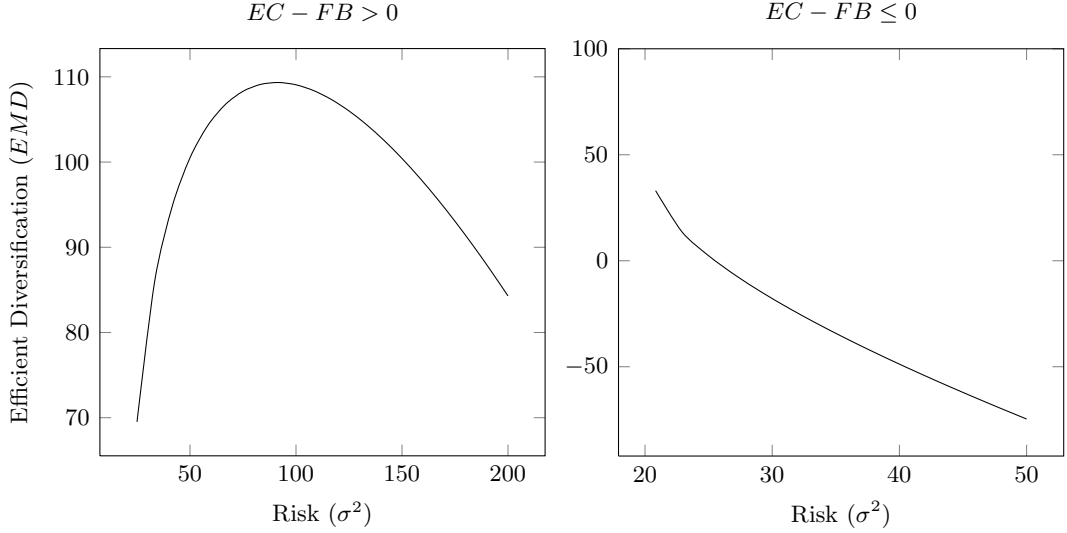
\begin{figure}[!t]
\caption{Risk and diversification relationship when investors have full information and asset returns distribution is not exchangeable: case where assets are risky}
\label{f3}
\begin{center}
\begin{tikzpicture}[]
\begin{axis}[my personal style,title=$EC-FB>0$,
xlabel=Risk ($\sigma^2$),
ymax=,
ymin=,
xmin=,
ylabel=Efficient Diversification ($EMD$),legend pos=north west,legend cell align=left,legend style={draw=none}]
\addplot+[mark=none,domain=0:200,smooth,black] {9.670447*sqrt((0.0483234*x-1)/0.01606466)-x+ 2.885208/0.0483234};
\end{axis}
\end{tikzpicture}
\begin{tikzpicture}[]
\begin{axis}[my personal style,title=$EC-FB\leq0$,
xlabel=Risk ($\sigma^2$),
ymax=100,
ymin=,
xmin=,
ylabel=,legend pos=north west,legend cell align=left,legend style={draw=none}]
\addplot+[mark=none,domain=0:50,smooth,black] {-0.7829712*sqrt((0.0483234*x-1)/0.0001219694)-x+ 2.885208/0.0483234};
\end{axis}
\end{tikzpicture}
\begin{minipage}{12cm}%
  \begin{spacing}{0.5}
    \footnotesize
We consider a universe of four assets. The variance-covariances matrix is $\boldsymbol{\Sigma}=\begin{pmatrix} 
185.0 & 86.5 &  80 & 20.0\\
 86.5 &196.0  & 76 & 13.5\\
80.0 & 76.0 & 411 &-19.0
\\
20.0 & 13.5 & -19 & 25.0
\end{pmatrix}$. In the case where $EC-FB>0$, $\boldsymbol{\mu}=\begin{pmatrix} 
14& 12& 15& 7
\end{pmatrix}^{\top}$. Otherwise i.e. $EC-FB\leq0$ $\boldsymbol{\mu}=\begin{pmatrix} 
0.14& 0.12& 0.15& 0.7
\end{pmatrix}^{\top}$.
\end{spacing}
  \end{minipage}
\end{center}
\end{figure}

\subsection{Risky free asset is available}

From \ref{eq16}, we have
\begin{equation}
1-w_f=\frac{\sigma\left(\boldsymbol{w}^{MV}\right)}{\sigma\left(\boldsymbol{w}^{tg}\right)}.\label{a17}
\end{equation}
Substituting $1-w_f$ from \ref{a17} into \ref{eq27}, we write the equation for the diversification as a function of risk as 
\begin{equation}
EDM\left(\boldsymbol{w}^{MV}\right)=\frac{\sum_{i=1}^{N}\boldsymbol{w}^{tg}_i\,\sigma_i^2}{\sigma\left(\boldsymbol{w}^{tg}\right)}\sigma\left(\boldsymbol{w}^{MV}\right)-\sigma^2\left(\boldsymbol{w}^{MV}\right).\label{a18}
\end{equation}
Substituting $\boldsymbol{w}^{tg}$ from \ref{eq14} and $\sigma\left(\boldsymbol{w}^{tg}\right)$ from \ref{eq15} into \ref{a18}, we have
\begin{equation}
EDM\left(\boldsymbol{w}^{MV}\right)=-\sigma^2\left(\boldsymbol{w}^{MV}\right)+\left(\frac{E-F\,\mu_f}{S}\right)\,\sqrt{\sigma^2\left(\boldsymbol{w}^{MV}\right)}.\label{a21}
\end{equation}
\cref{a21} represents the relationship between risk and diversification in the MV model when the risk free asset is available, short sales are allowed, investors hold the MV optimal portfolio and asset returns distribution is not exchangeable. The first partial derivative of $EDM\left(\boldsymbol{w}^{MV}\right)$ with respect of $\sigma^2\left(\boldsymbol{w}^{MV}\right)$ is
\begin{equation}
\frac{\partial EDM\left(\boldsymbol{w}^{MV}\right)}{\partial \sigma^2\left(\boldsymbol{w}^{MV}\right)}=-1+\left(\frac{E-F\,\mu_f}{2S}\right)\frac{1}{\sqrt{\sigma^2\left(\boldsymbol{w}^{MV}\right)}}.
\end{equation}
Its second partial derivative is
\begin{equation}
\frac{\partial^2 EDM\left(\boldsymbol{w}^{MV}\right)}{\partial \sigma^2\left(\boldsymbol{w}^{MV}\right)^2}=-\left(\frac{E-F\,\mu_f}{4S}\right)\left(\sigma^2\left(\boldsymbol{w}^{MV}\right)\right)^{-\frac{3}{2}}.
\end{equation} 
There are two cases in point depending of sign of $E-F\,\mu_f$, since $S>0$. Assume that $E-F\,\mu_f$ has negative sign i.e.  $E-F\,\mu_f\leq 0$. In that case,
\begin{align}
\frac{\partial EDM\left(\boldsymbol{w}^{MV}\right)}{\partial \sigma^2\left(\boldsymbol{w}^{MV}\right)}&<0\\
\frac{\partial EDM\left(\boldsymbol{w}^{MV}\right)}{\partial \sigma^2\left(\boldsymbol{w}^{MV}\right)^2}&>0
\end{align}
It follows that diversification is an decreasing convex function of risk. 

Now, assume that $E-F\,\mu_f$ has positive sign i.e. $E-F\,\mu_f\geq 0$. In that case, 
\begin{equation}
\begin{cases} \frac{ \partial EDM\left(\boldsymbol{w}^{MV}\right)}{\partial \sigma^2\left(\boldsymbol{w}^{MV}\right)}\leq 0 &\mbox{if } 0\leq\sigma^2\left(\boldsymbol{w}^{MV}\right)\leq \frac{(E-F\,\mu_f)^2}{4S^2}\\ 
\frac{ \partial EDM\left(\boldsymbol{w}^{MV}\right)}{\partial \sigma^2\left(\boldsymbol{w}^{MV}\right)}\geq 0 &\mbox{if } \sigma^2\left(\boldsymbol{w}^{MV}\right)\geq \frac{(E-F\,\mu_f)^2}{4S^2} \end{cases}
\end{equation}
and 
\begin{equation}
\frac{\partial EDM\left(\boldsymbol{w}^{MV}\right)}{\partial \sigma^2\left(\boldsymbol{w}^{MV}\right)^2}<0.
\end{equation}
It follows that diversification is an inverted U-shaped concave function of risk with maximum when risk is equal to $\frac{(E-F\,\mu_f)^2}{4S^2}$. 

\pgfplotsset{my personal style/.style=
{font=\footnotesize},width=7.5cm,height=7cm}
\begin{figure}[!t]
\caption{Risk and diversification relationship when investors have full information and asset returns distribution is not exchangeable: case where assets are risky}
\label{f4}
\begin{center}
\begin{tikzpicture}[]
\begin{axis}[my personal style,title=$E-F\,\mu_f>0$,
xlabel=Risk ($\sigma^2$),
ymax=,
ymin=,
xmin=,
ylabel=Efficient Diversification ($EMD$),legend pos=north west,legend cell align=left,legend style={draw=none}]
\addplot+[mark=none,domain=0:300,smooth,black] {-x+21.15671*sqrt(x)};
\end{axis}
\end{tikzpicture}
\begin{tikzpicture}[]
\begin{axis}[my personal style,title=$E-F\,\mu_f\leq0$,
xlabel=Risk ($\sigma^2$),
ymax=,
ymin=,
xmin=,
ylabel=,legend pos=north west,legend cell align=left,legend style={draw=none}]
\addplot+[mark=none,domain=0:10,smooth,black] {-x-4.470862*sqrt(x)};
\end{axis}
\end{tikzpicture}
\begin{minipage}{12cm}%
  \begin{spacing}{0.5}
    \footnotesize
We consider a universe of four assets. The variance-covariances matrix is $\boldsymbol{\Sigma}=\begin{pmatrix} 
185.0 & 86.5 &  80 & 20.0\\
 86.5 &196.0  & 76 & 13.5\\
80.0 & 76.0 & 411 &-19.0
\\
20.0 & 13.5 & -19 & 25.0
\end{pmatrix}$ and $\boldsymbol{\mu}=\begin{pmatrix} 
14& 12& 15& 7
\end{pmatrix}^{\top}$ and the risk free asset rate is $\mu_f=6$. In the case where $E-F\,\mu_f>0$, the risk free asset rate is $\mu_f=6$. Otherwise i.e. $E-F\,\mu_f\leq0$, the risk free asset rate is $\mu_f=13$.
\end{spacing}
  \end{minipage}
\end{center}
\end{figure}
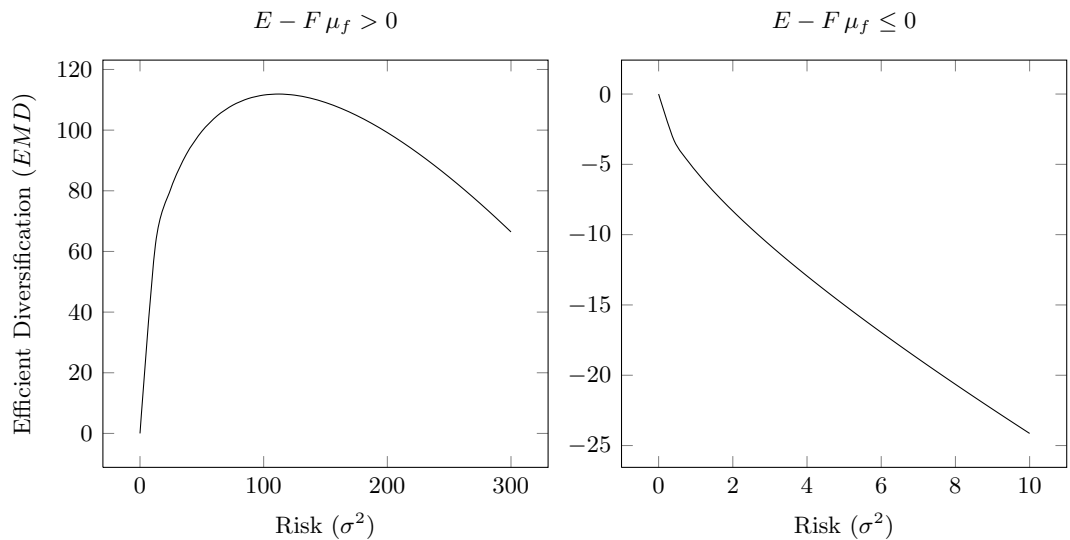
%
%
\newpage
\bibliographystyle{ecta}
\bibliography{rqediv_sm}

\end{document}